\documentclass{optica-article}

\journal{opticajournal} 

\articletype{Research Article}

\usepackage{lineno}

\begin{document}

\title{Offset sideband locking to iodine for laser cooling  on the $\mathrm{{}^1 S_0}\rightarrow \mathrm{{}^3 P_1}$ transition of ${}^{171}\mathrm{Yb}$}

\author{Martin Hauden,\authormark{1,*} Jacques Millo,\authormark{2} Martina Matusko,\authormark{1} Francisco S. Ponciano Ojeda, \authormark{1}  Yann Kersalé, \authormark{2} and Marion Delehaye\authormark{1}}

\address{\authormark{1}Université de Franche-Comté, SUPMICROTECH-ENSMM, CNRS, Institut FEMTO-ST, F-25000 Besançon, France\\
\authormark{2}SUPMICROTECH-ENSMM, Université de Franche-Comté, CNRS, Institut FEMTO-ST, F-25000 Besançon, France}
\email{\authormark{*}martin.hauden@femto-st.fr}

\begin{abstract}
We present absolute frequency measurements of a laser stabilized using an offset sideband locking technique on the P(49)24-1 rovibrational transition of $^{127}\mathrm{I}_2$ near 556 nm. The P(49)24-1 transition is offset by 4.8 GHz from the intercombination transition of ${}^{171}\mathrm{Yb}$. A dual-tone electro-optical modulator is employed to bridge this frequency gap, enabling frequency stabilization of the 556 nm laser near the Yb transition, achieving a fractional frequency stability of $1.2 \times 10^{-11}$ at one second. We thoroughly characterize the frequency shifts associated with this locking scheme, with typical systematic effects fractional uncertainties of $9 \times 10^{-12}$.
\end{abstract}

\section{Introduction}
The ytterbium (Yb) atom is widely used in quantum technologies, including quantum computing, precision measurements, and fundamental physics tests~\cite{yb_qbit_2022,yb_qbit_2022_2,yb_sun,yb_sun_2,yb_metrology,yb_metrology_2,yb_squeezing}. Laser-cooled Yb-based experiments typically use the 181~kHz-wide ${}^1\mathrm{S}_0-{ }^3\mathrm{P}_1$ intercombination transition at 556~nm, for example, to create a magneto-optical trap (MOT) with a Doppler temperature of 4.4~µK.

Standard frequency stabilization using sub-Doppler spectroscopy in a hot vapor cell is impractical for ytterbium due to its low saturation vapor pressure at room temperature. Alternative methods, such as direct spectroscopy of an atomic beam~\cite{dualMOT2016} or stabilization to an inverted crossover resonance~\cite{salter2017inverted}, offer potential solutions, but they are limited by moderate signal-to-noise ratios.

The P(49)24-1 iodine transition is particularly intense and is located only 1~GHz away from the ${}^1\mathrm{S}_0-{ }^3\mathrm{P}_1$ transition in ${}^{174}\mathrm{Yb}$. This frequency gap can be easily bridged using acousto-optic modulators (AOMs), making this transition popular for laser stabilization in experiments with the bosonic ${}^{174}\mathrm{Yb}$ isotope, where fractional frequency stability can reach the $10^{-12}$ level at one second~\cite{de2024laser}. In contrast, this iodine transition is 4.8 GHz away from the ${}^1\mathrm{S}_0 (F=1/2) - { }^3\mathrm{P}_1 (F'=3/2)$ transition in ${}^{171}\mathrm{Yb}$, leading many groups using ${}^{171}\mathrm{Yb}$ to rely on ultra-stable cavities~\cite{milani2017multiple} or OFCs~\cite{kohno2009one} for frequency stabilization.

In this work, we present an offset sideband locking scheme that utilizes a 4.7 GHz electro-optical modulator (EOM). Fine-tuning of the MOT beam frequency is performed using a 125~MHz acousto-optic modulator. The -1 sideband of the EOM is used to perform spectroscopy in an iodine cell, so that the carrier is kept close to resonance with the ${}^1\mathrm{S}_0 (F=1/2) - { }^3\mathrm{P}_1 (F'=3/2)$ transition in ${}^{171}\mathrm{Yb}$ at $539.390 405 756 (70)~\mathrm{THz}$\cite{McFerran_Yb_absolute}. Offset sideband locking to optical resonances has been previously demonstrated using an ultra-stable Fabry-Pérot cavity~\cite{thorpe2008laser, rabga2023implementing} and a Rb cell as references~\cite{peng2014locking}.

We perform a comprehensive analysis of the systematic frequency shifts introduced by this locking scheme and use the setup to measure the hyperfine transition frequencies of the P(49)24-1 iodine line at 556 nm. The measurements are referenced to an optical frequency comb (OFC), which is stabilized by both an ultra-stable laser and a hydrogen maser (HM) synchronized to UTC(OP). Our results are consistent with recent measurements~\cite{tanabe2022frequency}, with agreement at typically the 10 kHz level.

\section{Experimental setup}

\subsection{Sub-Doppler spectroscopy}
\begin{figure*}[h!]
\centering
\fbox{\includegraphics[width=1\linewidth]{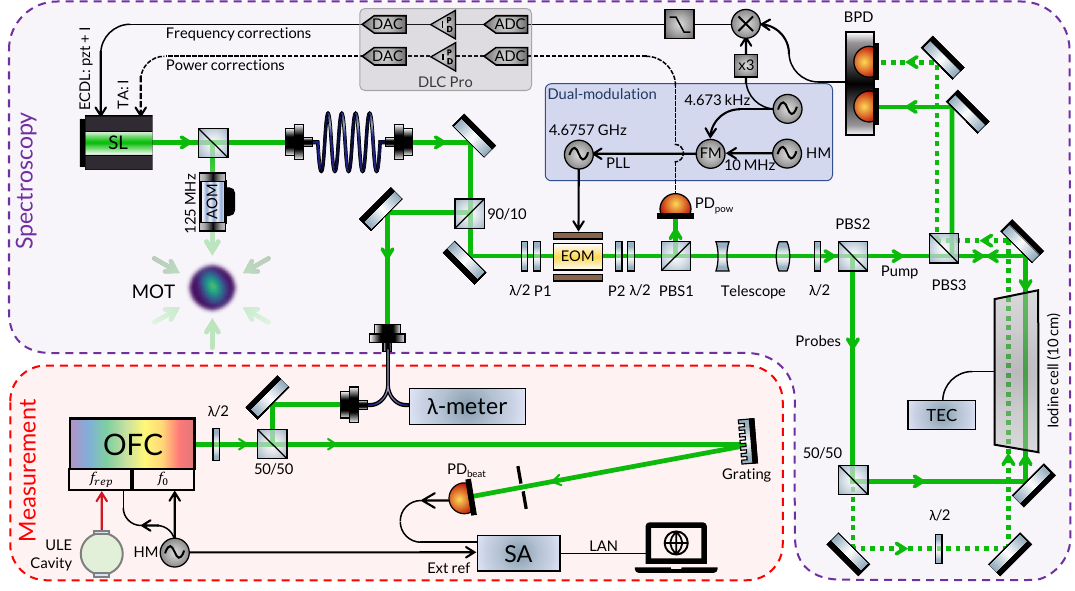}}
\caption{Detailed experimental setup used to lock the spectroscopy laser (SL) frequency to a 10$~$cm-long iodine vapor cell. Linear spectroscopy is shown in dotted line. Sub-Doppler spectroscopy is presented in solid lines. MOT: magneto-optical trap; P1/P2: polarizers; EOM: electro-optical modulator; PBS: polarizing beam splitter; PD: photodiode; BPD: balanced photodiode; HM: hydrogen maser; FM: frequency modulation; TEC: thermoelectric cooler; TA: tapered amplifier; OFC: optical frequency comb; SA: spectrum analyzer; ADC: analog to digital converter; DAC: digital to analog converter.}
\label{fig:setup}
\end{figure*}

The experimental setup for stabilizing our laser to the sub-Doppler spectroscopy of I$_2$ is shown in Fig. \ref{fig:setup}. The spectroscopy laser (SL) at 556~nm consists of an external cavity diode laser operating at 1112~nm, a tapered amplifier (TA), and a frequency-doubling cavity (Toptica TA-SHG pro system). The SL undergoes dual-frequency modulation using a resonant electro-optical modulator (EOM -- from Qubig GmbH). The first modulation is done with 0.5~W at $f_{\mathrm{EOM}} =  4.675700~\mathrm{GHz}$ and shifts the -1 order sideband into resonance with I$_2$, while the carrier is near resonance with the ${}^{171}\mathrm{Yb}$ ${}^1\mathrm{S}_0-{ }^3\mathrm{P}_1 (F'=3/2)$ transition. A second modulation at 4.673~kHz is applied for lock-in detection (LIA) by frequency modulating the 10~MHz external reference of the EOM driver. Residual amplitude modulation (RAM) in the EOM is passively minimized by aligning the light polarization with the crystal axes, which improves the SL frequency stability by an order of magnitude for integration times $\tau \geq 10^2~\mathrm{s}$. An optical power lock is implemented via the photodiode $\mathrm{PD{pow}}$, which adjusts the TA current, reducing power fluctuations to below 2\%.

Spectroscopy is performed in a 10$~$cm-long cell (Thorlabs GC19100-I), temperature-regulated at 10.1$^\circ$C. A pump-probe setup is used to generate a saturated absorption signal. To correct the Doppler background, a reference signal from a linear absorption setup using the same cell is subtracted with a balanced photodiode (BPD). The saturation intensity for this transition was measured to be $I_\mathrm{sat} = 2600 \pm 800 ~\mathrm{mW/cm^2}$ by fitting the linewidths at different intensities. The pump intensity is set to $I_\mathrm{sat}$, while the probes intensities are adjusted to $I_\mathrm{sat}/30$.

\begin{figure}[h!]
\centering
\fbox{\includegraphics[width=0.9\linewidth]{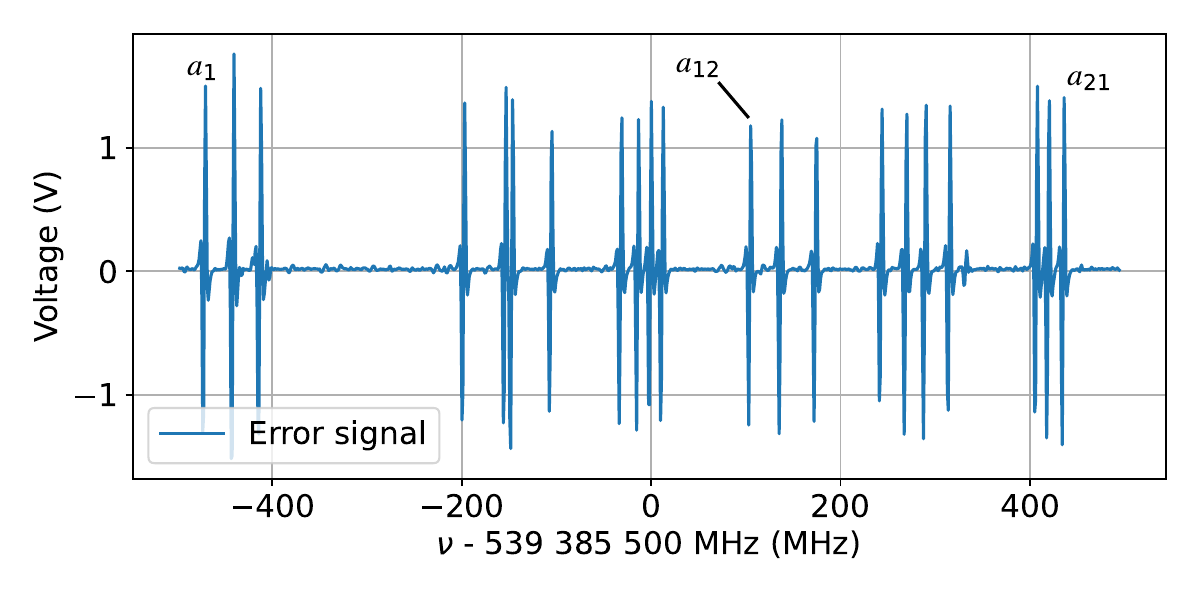}}
\caption{Error signal generated from lock-in detection. Transitions are named from $a_1$ to $a_{21}$. The $a_{12}$ line is used to lock the laser and address the ${}^1\mathrm{S}_0-{ }^3\mathrm{P}_1$ intercombination transition of Yb with an AOM. The signal-to-noise ratio is measured to be 100 with a bandwidth of 340~kHz.}
\label{fig:scan}
\end{figure}

Lock-in detection is performed using the third derivative to filter out the residual Doppler background in the spectrum, albeit with a reduction in signal intensity. The resulting error signal of our setup is shown in Fig.$~$\ref{fig:scan}. The hyperfine transitions of the P(49)24-1 line of ${ }^{127}$ I$_2$ are resolved and labeled from $a_1$ to $a_{21}$. 

\subsection{Frequency stabilization}
A digital PID controller is used to stabilize the laser frequency. Fast corrections are applied to the laser diode current (above 100 Hz), while slow corrections are applied to the piezoelectric element of the laser diode cavity. Imperfect RAM cancellation causes fluctuations in the error signal baseline, which degrades the frequency stability of the spectroscopy laser (SL). To mitigate this, we employ a cycled locking scheme where every 10 seconds, the laser is briefly unlocked, and a scan of the transition is performed. The baseline of the error signal is fitted and added as a compensation offset to suppress baseline drifts. After this adjustment, the laser is locked for the next cycle. It is possible to synchronize the cycle lock with the MOT loading sequence for optimum long-term laser stabilization without active RAM stabilization. This improved stabilization is however not necessary for a standard MOT operation.

\subsection{Frequency measurements}

The absolute frequency of the SL is measured by generating a beatnote with an ultra-stable reference, consisting of a 1542 nm laser frequency-locked to an ultra-stable Fabry-Perot cavity. Frequency stability and accuracy are transferred to the visible range using an optical frequency comb (OFC), which generates a series of discrete, equally spaced frequency lines (teeth), characterized by a repetition rate of $f_{\mathrm{rep}} \approx 250$ MHz and an offset frequency of $f_0 = 140$ MHz. The repetition rate is stabilized via a phase-locked loop to an ultra-low expansion glass cavity-stabilized laser~\cite{Didier2015}, frequency dedrifted by an active hydrogen maser. The offset frequency is locked using standard self-referencing techniques~\cite{telle1999carrier}. The fractional frequency stability of the OFC has been measured to be below $2 \times 10^{-14}$ from 1 s to $2 \times 10^4$ s, as shown in Fig.~\ref{fig:stab}. Since the OFC is located in a different experimental room, the SL signal is transmitted through a 50~m long fiber, which introduces fractional frequency noise below the $10^{-15}$ level\cite{matusko2023fully}.

The generation and measurement of the beatnotes between the SL and the OFC are illustrated in Fig.~\ref{fig:setup}. The interferences of the OFC and SL produce beatnotes at frequencies $f_b^+$ and $f_b^-$, corresponding to the teeth $N$ and $N+1$, respectively. The resulting radio-frequency (RF) spectrum of these beatnotes is measured using a fibered fast photodiode $\mathrm{PD_{beat}}$. A grating with a dispersion of 0.8~nm/mrad is employed to spatially filter out undesired optical frequencies. The OFC power on the photodiode is less than 0.5 µW per tooth, while the SL power is set around 600 µW, resulting in a detected beatnote power of -75~dBm. This power level is approximately 20~dB above the noise floor measured with a spectrum analyzer (SA -- R\&S FPC1000, span 10~MHz, resolution bandwidth 100~kHz, and video bandwidth 10~kHz).

The SL frequency $\nu_l$ can be expressed in terms of the OFC parameters and the measured beatnote frequency $f_{b}^\pm$  as:
\begin{eqnarray}
\nu_{l} &=& f_{0} + Nf_{rep} + f_{b}^+ \\
         &=& f_{0} + (N+1)f_{rep} - f_{b}^-, \notag
\end{eqnarray}
where $N \approx 2~157~000$ is determined  with a 10~MHz absolute accuracy wavemeter (HighFinesse WS8-2 calibrated on a He-Ne laser).

The limited signal-to-noise ratio of the beatnote hinders the use of a frequency counter or a tracking oscillator, even with amplification and filtering. Consequently, we employed an alternative method using a SA referenced to a hydrogen maser. We optimized the SA parameters to achieve the best frequency resolution for each scan while keeping the sweep time under 30 ms. Data is averaged over 10 scans before collection and fitting. A Lorentzian function is used to extract the central frequency once per second. Given the intermittent nature of the frequency read-out and the sweep of the measurement window across the SA span, this method likely overestimates the instability.
To ensure that the measurement scheme is not a limiting factor, we performed the following test: an RF synthesizer, referenced by another independent HM, generates a -80~dBm signal for the SA, which is analyzed as previously described. The fractional frequency stability transferred to optical frequencies was measured to be $5.5 \times 10^{-13} \tau^{-1/2}$, which is more than an order of magnitude better than the stability of the spectroscopy laser (SL), as shown in Fig.~\ref{fig:stab}. 
The SA accuracy is measured to be below 0.4~kHz by delivering the non-attenuated signal to a frequency counter during the test. 
Extraction of central frequencies from a Lorentzian fit of a power spectral density has been mentioned by another group for frequency stability measurements at the $10^{-15}$ level~\cite{norcia}.

\section{Results}

\subsection{Frequency stability}

\begin{figure}[h!]
\centering
\fbox{\includegraphics[width=0.98\linewidth]{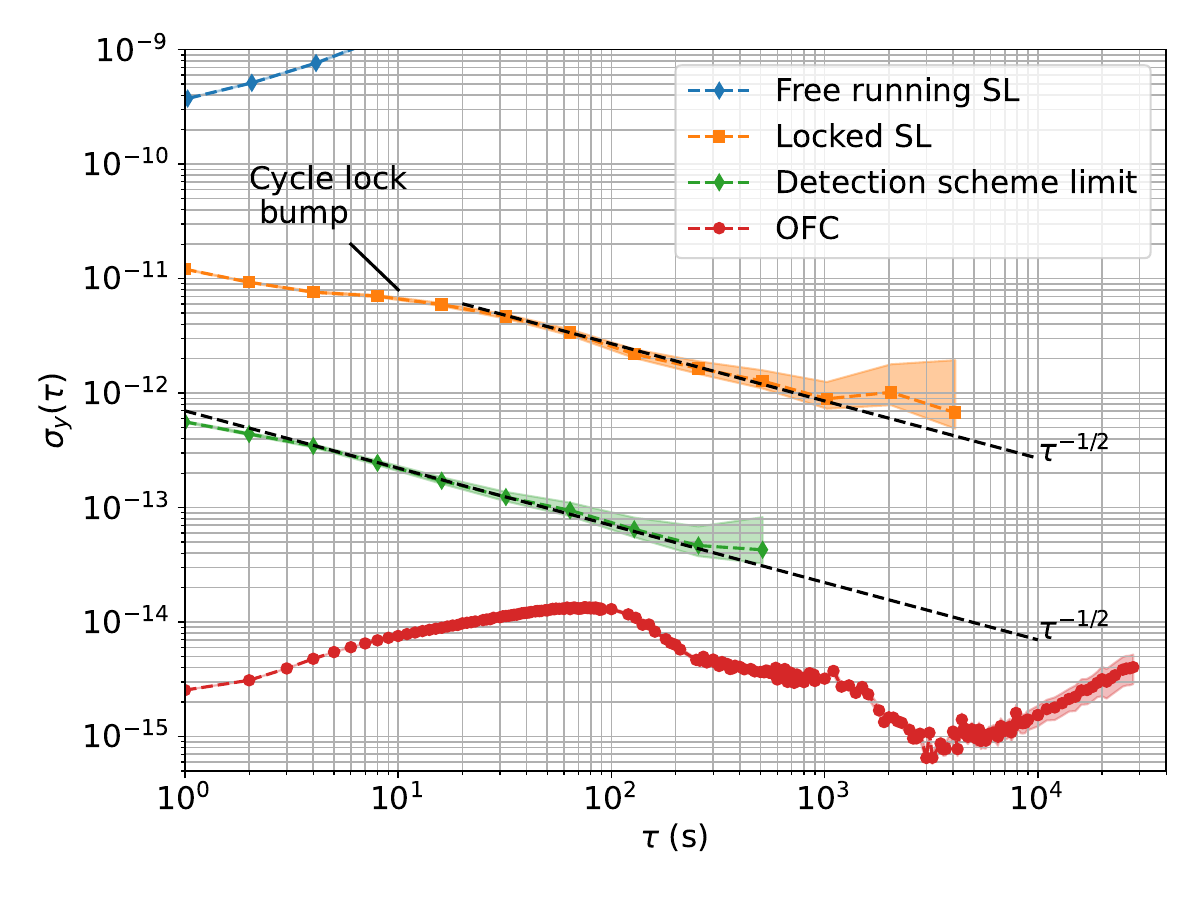}}
\caption{In blue, fractional frequency instabilities of the free running spectroscopy laser (SL). In orange, instability of a typical absolute frequency measurement of the SL. In green, detection limit of the SA brought to optical frequencies. In red, OFC instability with drift reduction. Dashed lines are guides for the eye.}
\label{fig:stab}
\end{figure}

Frequency stabilization of the SL to molecular iodine results in fractional frequency stability in the low $10^{-11}$ range at one second, representing an improvement of more than an order of magnitude over the free-running regime. The stability integrates as $\tau^{-1/2}$ from 10~s down to $8\times10^{-13}$ at 1000~s as shown in Fig.$~$\ref{fig:stab}.  This performance is comparable at long term (1000~s) to what has been reported using modulation transfer spectroscopy in \cite{tanabe2022frequency,de2024laser}. However, the cycling lock introduces instabilities at $\tau = 10$ seconds. Overall, this stability is suitable for precision measurements of the ${}^1\mathrm{S}_0-{ }^3\mathrm{P}_1$ intercombination transition of ytterbium during extended operations, with typical integration times of around  2000$~$s.

\subsection{Frequency shifts}
The measurement of I$_2$ line frequencies is subject to various types of shifts, which can be categorized into atomic transition shifts and measurement-induced shifts. These shifts are detailed in  Table~\ref{tab:fshifts}.

\begin{table}[htbp]
\centering
\caption{\bf Table of the frequency shifts present in the experimental setup and the corresponding maximum uncertainties. Statistical uncertainty refers to the stability of the SL.}
\begin{tabular}{ c c c } 
 
 Effect & Shift & Uncertainty  \\ 
 \hline
 Pressure shift & $(-7.58 \pm 0.07)~\mathrm{kHz}/\mathrm{Pa}\times P$ & $ 0.4$ kHz \\ 
 
 Asymmetry shift & $(-224 \pm 5)~\mathrm{kHz} \times a$ & $ 2$ kHz \\ 
 
 Residual light shift & $(-141 \pm 73)~\mathrm{Hz}/\mathrm{mW} \times \mathrm{power}$& $ 0.1$ kHz \\ 

 Alignment shift & $(6 \pm 4)~\mathrm{kHz}/\mathrm{^\circ}\times\mathrm{angle}$ & $ 0.4$ kHz \\ 
 
 HM shift & $610~\mathrm{Hz}$ &  $ 6$ Hz\\ 
 
 Magnetic shift & $100~\mathrm{Hz}$ & $ 0.1$ kHz \\ 
 
 EOM & $-4.675700~\mathrm{GHz}$ &  \\
 
 Cell impurity & &$ 5$~kHz \\
 
 Gravitational shift & 15 Hz & $ 1$ Hz\\
 \hline
 Statistical & & $ 2.1$ kHz \\
 Fit error  & & $ 0.4$ kHz \\
 \hline
 Total & & $ 5.9$ kHz
\end{tabular}
  \label{tab:fshifts}
\end{table}

\begin{figure*}[h!]
\centering
\fbox{\includegraphics[width=0.95\linewidth]{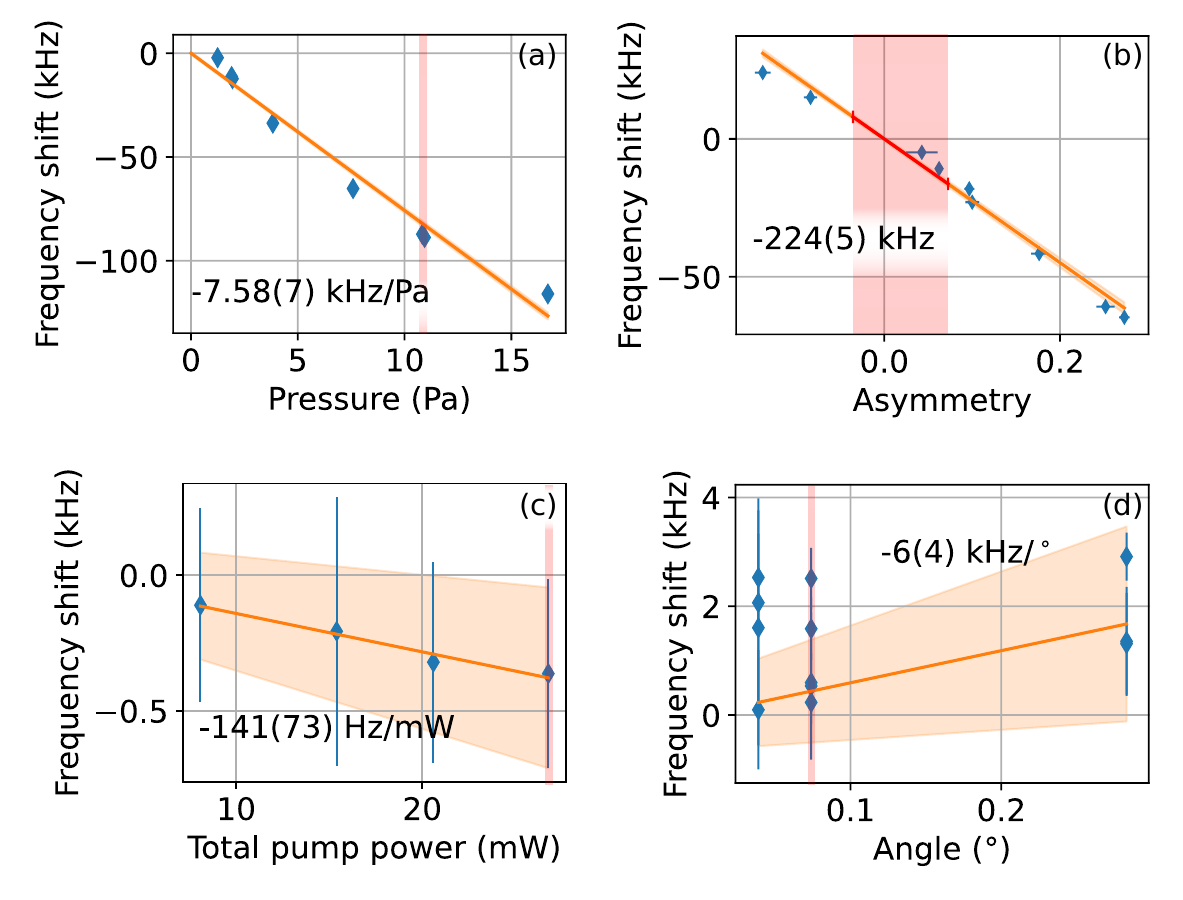}}
\caption{In blue: data points. In orange: linear fit. Frequency sensitivity to: a) Cell pressure; b) Error signal asymmetry; c) Residual light shift; d) Misalignment induced shift. Error bars are given for $1~\sigma$. Orange areas: fit uncertainties. Red areas: operating point(s).}
\label{fig:sensi}
\end{figure*}

\textbf{Pressure shift} is investigated by varying the iodine cell cold finger temperature. The vapor pressure  is determined using an empirical formula~\cite{gillespie1936normal}. Temperature gradients may affect the reliability of the cold finger temperature, and consequently the iodine pressure. We checked that the logarithm of transmitted fraction of the unmodulated laser intensity was proportional to the estimated pressure in agreement with the Beer-Lambert and the ideal gas laws. The correlation coefficient $R^2 = 0.97$ for the pressure range of 0.9 - 20~Pa confirms the accuracy of our temperature measurements. Above 20~Pa, we observe deviations to Beer's law that we attribute to limited iodine density in the cell. The pressure shift is measured on the  $a_{12}$ line for pressures between 1.3 and 20~Pa to be  $-7.58 \pm 0.02~\mathrm{kHz/Pa}$ [see Fig.~\ref{fig:sensi} (a)]. The slope measured by~\cite{tanabe2022frequency} is $-1.8~\mathrm{kHz/Pa}$, however both extrapolated zero-pressure measurements agree within the error bars (see discussion \ref{sec:disc}). 

\textbf{Asymmetry} in the error signal has been observed, causing a shift in the zero crossing from the unperturbed transition. 
We model the asymmetry by assuming a frequency-dependent linewidth $\gamma$~\cite{stancik2008simple}:

\begin{equation}
\gamma(\nu)=\frac{2 \gamma_0}{1+\exp \left[\frac{a}{\gamma_0}\left(\nu-\nu_0\right)\right]},
\label{eq:asy}
\end{equation}
where $2\gamma_0 = 7~\textrm{MHz}$ is the power-broadened linewidth of the transition, $\nu_0$ the resonance frequency and $a$ the dimensionless asymmetry parameter. The line is well described by an asymmetric Voigt profile. However to routinely extract the asymmetry at each locking cycle, we rather inject Equation (\ref{eq:asy}) into only a simple Lorentzian expression.  This simplification does not compromise the consistency of our results and ensure reliable convergence of the fit.
The frequency shift of the transition varies linearly with $a$, with a coefficient  $\delta\nu/a = (-224 \pm 5)~\mathrm{kHz}$ [see Fig.~\ref{fig:sensi} (b)].

In our system, the dual-tone spectroscopy is the primary source of asymmetry. First,  distortion of the modulated RF signal by the resonant circuit of the EOM introduces amplitude modulation, leading to asymmetry~\cite{janik1986two, kauranen1994determination, avetisov1996two}. Second, non-resonant spectral components (including the carrier and other sidebands of the EOM) cause molecular line distorsion and light shifts~\cite{yudin2023frequency, hilton2020light}.
We have verified that electronic distortion from the LIA filters is negligible with respect to other sources of  distortion. Additionally, we confirmed that no significant asymmetry is caused by residual Doppler background, misalignment, RAM due to imperfect light polarization in the EOM~\cite{du2004correction}, or radiation pressure~\cite{grimm1989effect}.

\begin{figure}[h!]
\centering
\fbox{\includegraphics[width=0.95\linewidth]{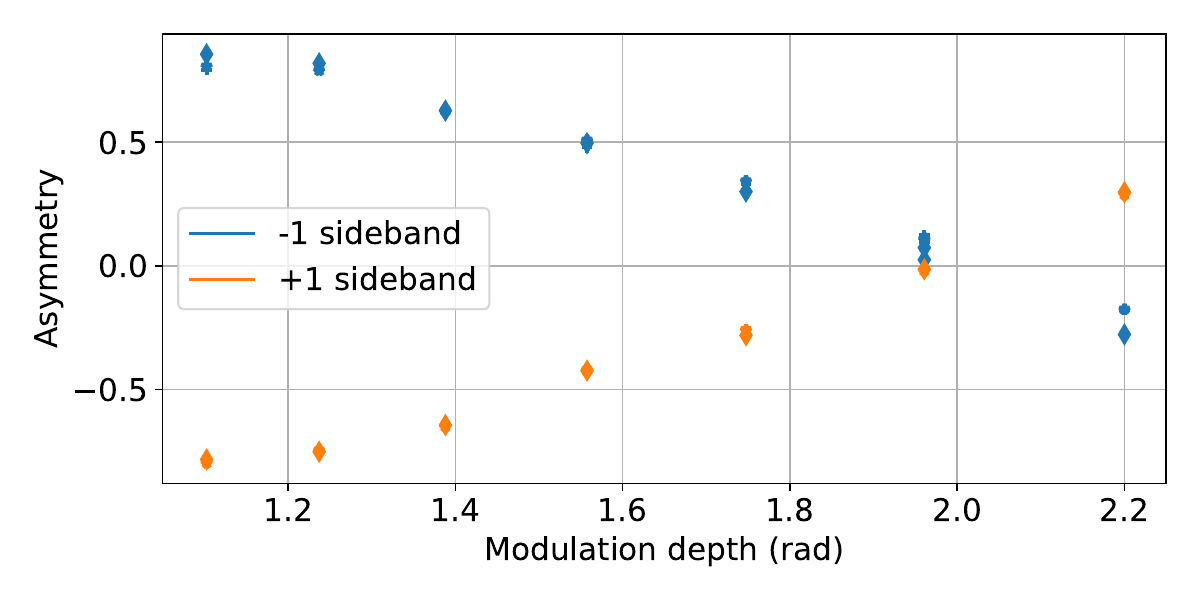}}
\caption{Asymmetry of the error signal measured as a function of the EOM modulation depth. In blue, the spectroscopy is performed with the $-1$ sideband and in orange with the $+1$. Stars refers to $a_{1}$, diamonds to $a_{12}$ and crosses to $a_{21}$ component of the transition.}
\label{fig:asy}
\end{figure}

We conducted asymmetry measurements by varying  the modulation depth of the EOM and using either the $-1$ or the $+1$ order sideband [see Fig.~\ref{fig:asy}].  This procedure revealed exactly opposite asymmetries for the two sidebands. By adjusting the EOM modulation depth to minimize asymmetry, we ensured that both sidebands yield the same absolute frequency measurements, thereby validating our approach.

\textbf{Light shift} is also investigated independently by measuring the frequency as a function of the pump power in the resonant sideband, while correcting for the asymmetry. A linear fit is performed to determine the residual sensitivity to light shifts, yielding a value of $(-141 \pm 73)~\mathrm{Hz/mW}$ for a beam diameter of 1.2~mm  [see Fig.~\ref{fig:sensi} (c)]. This value is significantly lower than the few kHz/mW only reference reported in the literature for this transition\cite{tanabe2022frequency}. Most of the light shift is effectively compensated by the asymmetry correction.

\textbf{Misalignment shift} is measured for different angles $\theta$ between the pump and probe wave vectors. Absolute frequency of iodine lines has been measured with an angle of $\theta = 0.07^\circ$, while additional measurements are taken for a smaller misalignment ($\theta = 0.04^\circ$) and a strongly misaligned case ($\theta = 0.3^\circ$). Several measurements are performed for each $\theta$ since the observed shifts are within our measurement uncertainty. A linear fit is used to determine the sensitivity to misalignment, which is found to be $6 \pm 4~\mathrm{kHz}/^\circ$ [see Fig.~\ref{fig:sensi} (d)]. The combination of the 7~MHz linewidth of the transition and the 1.2 mm-diameter beams leads to a low sensitivity to misalignment \cite{park2001dispersion}, that can be neglected here.

\textbf{Magnetic field shift} is measured to be $\lesssim 500~\textrm{Hz/G}$, comparable to a shift of $1062\pm6~\mathrm{Hz/G}$ that has been reported  at 514~nm~\cite{gillot2024influence}. This corresponds to a final shift of 0.1~kHz due to the residual magnetic field of $200$~mG near the cell.

\textbf{Hydrogen maser shift}: The HM used to measure the repetition rate of the OFC  exhibits a small offset of  $11.59\pm0.08~$µHz on the 10~MHz signal, as determined by comparison with UTC(OP).  This results in a  $610\pm6~$Hz shift in the absolute frequency measurements, which remains stable throughout the measurement campaign.

\textbf{Cell impurity}: Our cell was baked and evacuated to a pressure of $10^{-6}~\mathrm{Pa}$ by the manufacturer, while literature reports typical frequency shifts of a few kHz for impurity pressures around 1~Pa~\cite{fredin1989study,hrabina2008methods}. Consequently, we only conservatively apply the 5 kHz uncertainty recommended by the BIPM for frequency standards at 474 THz~\cite{bipm_webpage}.

\textbf{Gravitational shift}: FEMTO-ST is located 300~m above sea level, which induces a negligible 15~Hz shift of the transition frequencies. 

\subsection{Iodine P(49)24-1 transition referencing}
\label{sec:disc}
Molecular iodine $\mathrm{I_2}$ is a promising candidate  as a stable reference across the 500$~$nm to 800$~$nm range. It has been extensively used for laser frequency stabilization within its absorption spectrum. Laser stabilization using iodine has demonstrated excellent frequency stability, particularly at 532~nm \cite{arie1992absolute, hall1999stabilization, hong2004frequency, schuldt2021optical} and 515~nm \cite{borde1981high, wallerand2006absolute, ikeda2020iodine} due to the high contrast of the transitions. 
Considerable work has been dedicated to providing accurate frequency measurements of many transitions in iodine absorption spectrum. The atlas of iodine references transitions from 500$~$nm to 676$~$nm with accuracies ranging from MHz to tens of MHz \cite{gerstenkorn1978atlas}. These limited accuracies prompted further efforts to improve the referencing. Theoretical models and simulations have extended the covered range to 515 -- 830~nm~\cite{bodermann2002widely, knockel2004high}, with an accuracy dependent on existing experimental data at these wavelengths. Doppler-free spectroscopy has also been employed to enhance absolute frequency measurements of $\mathrm{I_2}$ over narrow frequency ranges at several wavelengths between 502 to 772~nm~\cite{du2005frequency,wallerand2006absolute,hong2000hyperfine,ma2006absolute,zhang2009absolute,yang2012hyperfine,reinhardt2006iodine,hong2009doppler,bernard2001absolute,edwards1996frequency,reinhardt2007absolute}.

During the elaboration of this work, an independent absolute frequency measurement of the P(49)24-1 transition was published, providing a basis for comparison with our results.
The SL was successively frequency-stabilized on each of the 21 components of the P(49)24-1 transition of molecular iodine. For each component, the laser frequency was measured at the kHz level and corrected using Table~\ref{tab:fshifts}. The measured frequencies for all 21 components are listed in Table~\ref{tab:fi2}, along with their associated uncertainties at the $1 \sigma$ level. Additionally, frequency differences between our measurements and those reported in \cite{tanabe2022frequency} for the same transition are presented. The mean frequency shift is measured to be -8~kHz, covered by the respective error bars. Residual systematic shifts are associated to uncertainty on the cold finger temperature.

\begin{table}[ht]
\centering
\caption{\bf Absolute laser frequency measurements ($f_{\mathrm{measured}}$), their associated correction ($f_c$), absolute iodine frequency, and frequency difference with \cite{tanabe2022frequency} for all hyperfine components of the P(49)24-1 line with their associated standard deviation.}
\begin{tabular}{ |c|c|c|c|c| } 
\hline
Line &$f_{\mathrm{measured}}$ (kHz)&$f_{\mathrm{c}}$ (kHz)& $f_{\mathrm{measured}}-f_{\mathrm{c}}-f_{\mathrm{EOM}}$ (kHz)  & $\Delta f$ with \cite{tanabe2022frequency} (kHz) \\ 
\hline
$\mathrm{a}_1$ & 539 389 707 845(1) & -91(5) & 539 385 032 236(6) & -3(9)\\ 
\hline
$\mathrm{a}_2$ & 539 389 736 697(2) & -89(5) & 539 385 061 086(6) & -1(9)\\
\hline
$\mathrm{a}_3$ & 539 389 763 996(2) & -87(6) & 539 385 088 383(6) & -24(9)\\ 
\hline
$\mathrm{a}_4$ & 539 389 977 543(1) & -89(5) & 539 385 301 932(5) & -4(9)\\ 
\hline
$\mathrm{a}_5$ & 539 390 021 113(1) & -86(5) & 539 385 345 499(5) & -18(9)\\ 
\hline
$\mathrm{a}_6$ & 539 390 028 090(1) & -87(5) & 539 385 352 477(5) & -2(9)\\ 
\hline
$\mathrm{a}_7$ & 539 390 069 683(1) & -87(5) & 539 385 394 070(6) &-4(9)\\ 
\hline
$\mathrm{a}_8$ & 539 390 143 394(2) & -89(6) & 539 385 467 783(6) & 1(9)\\ 
\hline
$\mathrm{a}_9$ & 539 390 161 201(1) & -81(5) & 539 385 485 582(5) & -6(9)\\ 
\hline
$\mathrm{a}_{10}$ & 539 390 174 760(3) & -80(5) & 539 385 499 140(6) & -4(9)\\ 
\hline
$\mathrm{a}_{11}$ & 539 390 187 147(1) & -84(5) & 539 385 511 531(5) & -10(9)\\ 
\hline
$\mathrm{a}_{12}$ & 539 390 279 716(1) & -94(5) & 539 385 604 110(5) & -1(9)\\ 
\hline
$\mathrm{a}_{13}$ & 539 390 312 238(2) & -86(5) & 539 385 636 624(6) & -5(9)\\ 
\hline
$\mathrm{a}_{14}$ & 539 390 348 924(1) & -87(5) & 539 385 673 311(5) & -20(9)\\ 
\hline
$\mathrm{a}_{15}$ & 539 390 418 492(1) & -93(6) & 539 385 742 885(6) & -3(9)\\ 
\hline
$\mathrm{a}_{16}$ & 539 390 444 614(1) & -85(5) & 539 385 768 999(6) & -6(9)\\ 
\hline
$\mathrm{a}_{17}$ & 539 390 465 353(2) & -86(6) & 539 385 789 739(6) & -6(9)\\ 
\hline
$\mathrm{a}_{18}$ & 539 390 491 278(1) & -83(5) & 539 385 815 661(6) & -9(9)\\ 
\hline
$\mathrm{a}_{19}$ & 539 390 585 259(2) & -89(5) & 539 385 909 648(6) & -2(9)\\ 
\hline
$\mathrm{a}_{20}$ & 539 390 597 983(1) & -87(5) & 539 385 922 370(6) & -21(9)\\ 
\hline
$\mathrm{a}_{21}$ & 539 390 614 296(1) & -69(5) & 539 385 938 665(5) & -22(9)\\ 
\hline
\end{tabular}
  \label{tab:fi2}
\end{table}

Following a procedure similar to that described in~\cite{tanabe2022frequency}, we use this data to extract the values of the constants of a hyperfine Hamiltonian including the electric quadrupole,  spin-rotation, tensor spin-spin and scalar spin-spin terms~\cite{borde1981high}. The standard deviation of the four-terms Hamiltonian fitting is 7~kHz and our values for the hyperfine parameters are consistent with those reported in \cite{tanabe2022frequency}, within the respective uncertainties, as shown in Table~\ref{tab:hf}.

\begin{table}[ht]
\centering
\caption{\bf Extracted values for hyperfine constants.}
\begin{tabular}{ |c|c|c| } 
\hline
Constant & this work (kHz) &  values from \cite{tanabe2022frequency} (kHz) \\
\hline
$\Delta eQq$ & 1 921 374(11) & 1 921 373.7(2.2) \\
\hline
$ \Delta C $ & 53.195(9) & 53.2255(26) \\
\hline
$\Delta d $ & -25.12(64) & -25.382(94) \\
\hline
$\Delta \delta $ & -7.29(53) & -6.589(55)\\
\hline

\hline
\end{tabular}
  \label{tab:hf}
\end{table}

This work demonstrates that offset sideband locking on iodine can be used to frequency-stabilize a laser with kilohertz-level accuracy and 1-second stability. This method is therefore suitable for stabilizing a laser to the resonance of the $\mathrm{{}^1 S_0}\rightarrow \mathrm{{}^3 P_1}$ transition in ${}^{171}\mathrm{Yb}$. Using this technique, we successfully operated a MOT for ${}^{171}\mathrm{Yb}$ with $4\times 10^7$ atoms at a temperature below 100~$\upmu \mathrm{K}$.

\section{Conclusion}

Our study presents a robust laser frequency stabilization scheme using an iodine vapor cell, suitable for addressing the ${}^1\mathrm{S}_0-{ }^3\mathrm{P}_1$ intercombination transition of $^{171}$Yb with a dual-tone EOM. Sub-Doppler spectroscopy techniques and lock-in techniques provide an effective error signal for frequency corrections.

Calibration and validation of the experimental setup were performed using frequency references available at FEMTO-ST, which were transferred to the SL frequency through an OFC. Our results demonstrate fractional frequency instabilities in the low $10^{-11}$ range at a one-second. We further improved long-term frequency stability with a cycle locking scheme, achieving instabilities as low as $8 \times 10^{-13}$ after one hour of integration. Additionally,  we qualify a measurement scheme designed for low-power optical beat notes utilizing an RF spectrum analyzer for stability and accuracy measurements.

Finally, we characterized frequency shifts with a focus on line asymmetry and confirmed absolute frequency measurements of the the hyperfine components of the P(49)24-1 rovibrational transition of $^{127}$I$_2$ near 556~nm at the 10~kHz level. This scheme is therefore  well-suited  for laser stabilization on relatively narrow atomic transitions.

\section{Acknowledgements}
This work has been supported by the EIPHI Graduate school (contract ANR-17-EURE-0002), the Bourgogne-Franche-Comté Region, the ANR (project CONSULA, ANR-21-CE47-0006-02), First-TF Labex (contract ANR-10-LABX-48-01) and Oscillator-IMP Equipex (contract ANR-11-EQPX-0033). The authors thank Clément Lacroute and Emmanuel Klinger for careful reading of the manuscript and Prof. Eberhard Tiemann for providing us transition frequency predictions near 556~nm. 

\section{Disclosures}
The authors declare that there are no conflicts of interest related to this article.
\bibliography{Biblio}

\end{document}